\documentclass[a4paper,11pt]{article}
\begin{document}
\def\baselinestretch{1.5}
{\def\title{Irreversibility and classical mechanics laws}
\def\author{V.M. Somsikov}
\def\correspondence{Prof. V.M. Somsikov}
\def\correspondenceshort{Prof. Somsikov}
\def\date{\today}
\def\addra{Laboratory of Physics of the geogeliocosmic relation}
\def\addrb{Institute of Ionosphere}
\def\addrc{Kamenskoe Plato}
\def\addrd{Almaty, 480020, Kazakhstan}
\def\tel{~~  ~~~+8-3272-548074~~~~~~~~~~~~~~~~~~~~~}
\def\fax{~~  ~~~+8-3272-658085~~~~~~~~~~~~~~~~~~~~~}
\def\email{~~  ~~ nes@kaznet.kz ~~~~~~~~~~}}
\small
\title {The irreversibility and classical mechanics laws}
\author{Vyacheslav Somsikov }
\date{\it{Laboratory of Physics of the geogeliocosmic relation,\\
Institute of Ionosphere, Almaty, 480020, Kazakhstan}} \maketitle
\begin{abstract}
The irreversibility of the dynamics of the conservative systems on
example of hard disks and potentially of interacting elements is
investigated in terms of laws of classical mechanics.  The
equation of the motion of interacting systems and the formula,
which expresses the entropy through the generalized forces, are
obtained. The explanation of irreversibility mechanism is
submitted. The intrinsic link between thermodynamics and classical
mechanics was analyzed.
\end{abstract}
\bigskip

\date{\it{Introduction.}} \maketitle

Irreversibility determines the contents of the second law of
thermodynamics in fundamental physics. According to this law there
is function S named entropy, which only grows for isolated
systems, achieving a maximum in an equilibrium state. But the
potentiality of forces for all four known fundamental interactions
of elementary particles is a cause of reversibility of the Newton
equation [1] and therefore declares that the natural processes
also should be reversible. Thus, the fundamental physics includes
parts contradicting each other: there are reversible classical
mechanics and irreversible thermodynamics.

The first attempt to resolve these contradictions belongs to
Boltzmann. From the $H$-theorem it follows that the many-body
systems should equilibrates. But at reception of these results
Boltzmann used probabilistic principles. Therefore the
contradiction between classical mechanics and statistical physics
has not been overcome. Great importance of this problem for all
physics explains its big popularity among the famous physicists of
the world. The history of its solution is very extensive and at
times, dramatic. Therefore let us reference only several works,
which precisely enough describe a modern condition of a problem of
irreversibility [2-4].

For overcoming of the irreversibility problem I. Prigogine has
suggested to try to create the expanded formalism of open systems
within the framework of classical mechanics laws [4]. Indeed, by
analyzing dynamics properties of a hard disks system it turned out
that such formalism appears in a process of the solution of a
problem of irreversibility [5-9]. For example, as a result of this
analyzing the generalized Liouville equation and equation of the
motion for the systems was obtained. Here on the background of
results [5-9] the explanations of the mechanism of
irreversibility, interrelation of classical mechanics and
thermodynamics are offered. All this determines a way of
construction of the expanded formalism of open systems within the
framework of classical mechanics laws.

Our investigation is based on the next method. The conservative
system of interacting elements is prepare in the nonequilibrium
way. This system is splitting on subsystems so that they could be
accepted as in equilibrium. The subsystems dynamics under
condition of their interactions is analyzed.

First of all we investigate the dynamics of the simplest
hard-disks systems. For those we find that so-called generalized
forces between selected subsystems should depend from velocities
of those. This leads to the key idea about impossibility of
solution of the problem of irreversibility in terms of canonical
Hamilton or Liouville equations. So, we have built the generalized
Hamilton and Liouville equations. Then we deduce how the
irreversible dynamics follows from the generalized Liouville
equation.

After that we analyze the dynamics of Newtonian systems and deduce
the equations of the motion for the underlying subsystems. On the
base of these equations we answer the question: why and how the
velocity dependence for the generalized forces between the
subsystems appears for the case of the potential-type forces
between the elements. Next point: the intrinsic link between
thermodynamics and classical mechanics was analyzed. The general
formula for the entropy of the system in terms of generalized
forces is obtained which corresponds to irreversible
transformation of the interaction energy of subsystems into the
internal energy by means of the generalized forces work.

\date{\it{Irreversibility in a hard-disk system.}} \maketitle

We used an approach of pair interaction of disks. Their motion
equations are deduced with help of the matrix of pair collisions.
As complex plane this matrix is given [5]:

$S_{kj}=\left(\begin{array}{cc} a & -i b
\\ -i b & a\end{array} \right)$ (a),

where $a=d_{kj}\exp(i\vartheta_{kj})$; $b=\beta
\exp(i\vartheta_{kj})$; $d_{kj}=cos\vartheta_{kj}$;
$\beta=sin\vartheta_{kj}$; $i$- an imaginary unit; $k$- and $j$-
numbers of colliding disks; $d_{kj}$- the impact parameter (IP),
determined by distance between centers of colliding disks in the
Cartesian plane system of coordinates with axes of $x$ and $y$, in
which the $k$-disk swoops on the lying the $j$- disk along the $x$
- axis. The scattering angle $\vartheta_{kj}$ varies from $0$ to
$\pi$. In consequence of collision the transformation of disks
velocities can be presented in such form: $V_{kj}^{+}=S_{kj}
V_{kj}^{-}$ (a), where $V_{kj}^{-}$ and $V_{kj}^{+}$ - are
bivectors of velocities of $k$ and $j$ - disks before $(-)$, and
after $(+)$ collisions, correspondingly; $V_{kj}$=$\{V_k,V_j\}$,
${V}_j=V_{jx}+iV_{jy}$ - are complex velocities of the incident
disk and the disk - target with corresponding components to the
$x$- and $y$- axes. The collisions are considered to be central,
and friction is neglected. Masses and diameters of disks-$"d"$ are
accepted to be equal to $1$. Boundary conditions are given as
either periodical or in form of hard walls. From (a) we can obtain
equations for the change of velocities of colliding disks [5-7]:
\begin{equation}
{\left(\begin{array}{c} {\delta V_k}\\ {\delta V_j}
\end{array} \right )
=\varphi_{kj} \left( \begin{array}{c} \Delta_{kj}^{-} \\
-\Delta_{kj}^{-}\end{array} \right )}.\label{eqn1}
\end{equation}
Here, $\Delta_{kj}=V_k-V_j$ - are relative velocities, $\delta
V_k=V_k^{+}-V_k^{-}$, and $\delta V_j=V_j^{+}-V_j^{-}$ - are
changes of disks velocities in consequence of collisions,
$\varphi_{kj}=i\beta e^{i\vartheta_{kj}}$. \\That is, eq. (1) can
be presented in the differential form [5]:
\begin{equation}
\dot{V}_k=\Phi_{kj}\delta (\psi_{kj}(t))\Delta_{kj}\label{eqn2}
\end{equation}
where $\psi_{kj}=[1-|l_{kj}|]/|\Delta_{kj}|$;
$\delta(\psi_{kj})$-delta function; $l_{kj}(t)=z_{kj}^0+\int
\limits_{0}^{t}\Delta_{kj}{dt}$ - are distances between centers of
colliding disks; $z_{kj}^0=z_k^0-z_j^0$, $z_k^0$ and $ z_j^0$ -
are initial values of disks coordinates;
$\Phi_{kj}=i(l_{kj}\Delta_{kj})/(|l_{kj}||\Delta_{kj}|)$.

    The equation (2) determines a redistribution of kinetic energy
between the colliding disks. It is not a Newtonian equation
because the forces between the colliding disks depend from their
relative velocities. So, we get the generalized Hamilton equation
to be applied for studying the subsystems dynamics [6]:

\begin{equation}
{\frac{\partial{H_p}}{\partial{r_k}}=-\dot{p_k}+F_{k}^p}\label{eqn3}
\end{equation}
\begin{equation}
{\frac{\partial{H_p}}{\partial{p_k}}={V_k}}\label{eqn4}
\end{equation}
These are the general Hamilton equations for the selected
$p$-subsystem. The external forces, which acted on $p$-subsystem,
presented in a right-hand side an eq.(3).

    Using eq.(3,4), we can find the Liouville equation for
$p$-subsystem. For this purpose, let's to take a generalized
current vector - $J_p=(\dot{r_k},{\dot{p_k}})$ of the
$p$-subsystem in a phase space. From equations (3,4), we find
[6,7] :
\begin{equation}
{\frac{df_p}{dt}=-f_p\sum\limits_{k=1}^T
\frac{\partial}{\partial{p_k}}F_{k}^p} \label{eqn5}
\end{equation}

Eq.(5) is a Liouville equation for $p$-subsystem. It has a formal
solution:\\${f_p=const\cdot{\exp{[-\int\limits_{0}^{t}{(\sum\limits_{k=1}^T
\frac{\partial}{\partial{p_k}}F_{k}^p)}{dt}]}}}$.

The equation (5) is obtained from the common reasons. It is
suitable for any interaction forces of subsystems. Thus, the
eq.(5) is applicable to analyze any open nonequilibrium systems.
In particularly, it can be used for explanation of
irreversibility.

The right side of eq.(5), $f_p\sum\limits_{k=1}^T
\frac{\partial}{\partial{p_k}}F_{k}^p$, is the integral of
collisions. This integral can be obtained from the motion
equations of the systems element. For example, for a hard disks
system it can be found with the help of the eq. (2).

Let's consider the important interrelation between descriptions of
dynamics of separate subsystems and dynamics of system as a whole.
As the expression, ${\sum\limits_{p=1}^R{\sum\limits_{k=1}^T
F_{k}^p =0}}$, is carried out, the next equation for the full
system Lagrangian, $L_R$, will have a place:
\begin{equation}
{\frac{d}{dt}\frac{\partial{L_R}}{\partial{V_k}}-
\frac{\partial{L_R}}{\partial{r_k}}=0}\label{eqn6}
\end{equation}
and the appropriate Liouville equation:
${\frac{\partial{f_R}}{\partial{t}}+{V_k}\frac{\partial{f_R}}
{\partial{r_k}}+\dot{p_k}\frac{\partial{f_R}}{\partial{p_k}}=0}$.
The function, $f_R$, corresponds to the full system. The full
system is conservative. Therefore, we have: ${\sum\limits_{p=1}^R
divJ_p=0}$. This expression is equivalent to the next equality:
${\frac{d}{dt}(\sum\limits_{p=1}^{R}\ln{f_p})}=
\frac{d}{dt}(\ln{\prod\limits_{p=1}^{R}f_p})=
{(\prod\limits_{p=1}^{R}f_p)}^{-1}\frac{d}{dt}(\prod\limits_{p=1}^{R}{f_p})=0$.
So, $\prod\limits_{p=1}^R{f_p}=const$. In an equilibrium state we
have $\prod\limits_{p=1}^R{f_p}=f_R$. Because the equality
$\sum\limits_{p=1}^{R}F_p=0$ is fulfilled during all time, we have
that equality, $\prod\limits_{p=1}^R{f_p}=f_R$, is a motion
integral. It is in agreement with Liouville theorem about
conservation of phase space [10]. So, only in two cases the
Liouville equation for the whole system is in agreement with the
general Liouville equation for selected subsystems: if the
condition $\int\limits_{0}^{t}{(\sum\limits_{k=1}^T\frac{\partial}
{\partial{p_k}}F_{k}^p)}dt\rightarrow{const}$ (c) is satisfied
when $t\rightarrow\infty$, or when,
${(\sum\limits_{k=1}^T}\frac{\partial}{\partial{p_k}}F_{k}^p)$, is
a periodic function of time. The first case corresponds to the
irreversible dynamics, and the second case corresponds to
reversible dynamics. Because the generalized forces for a
hard-disk system depended from velocities, the irreversible
dynamics is possible.

Dynamics of strongly rarefied systems of potentially interacting
elements is also described by the eq. (2). Therefore for those
systems the irreversibility is possible as well [7,8].

The dependence of generalized force from velocities - it is a
necessary condition for the irreversibility was really to be
occurred. Therefore the question about irreversibility for
Newtonian systems is reduced to that about the velocity dependence
of forces between subsystems.

For hard disks and strongly rarefied systems of potentially
interacting elements the presence of irreversibility is
predetermined by equation of motion (2). In these systems the
interaction forces between the elements is depending from
velocities. Therefore it is clear that the generalized forces will
depend from velocities also. But forces between elements for
Newtonian systems are potentional. Therefore it is necessary to
answer the question: how velocity dependence of generalized force
is appearing when forces between the elements are independent from
velocities. For this purpose in the next part we will obtain of
the motion  equation for subsystems.

\date{\it{The evolutionary equation of the motion for subsystem.}}\maketitle

Let us to analyze Newtonian systems. We take a system with energy:
$E_N=T_N+U_N=const$, where
$T_N=\frac{1}{2}\sum\limits_{i=1}^N{{v_i}^2}$ -is a kinetic
energy; $U_N(r_{ij})$ - is a potential energy; $r_{ij}=r_i-r_j$ -
is a distance between $i$ and $j$ elements; $N$ - is a number of
elements. Masses of elements are accepted to $1$.

The equation of the motion for elements is:
\begin{equation}
{v_i=-\sum\limits_{i=1,j\neq{i}}^N\frac{\partial}
{\partial{r_{ij}}}U}\label{eqn7}
\end{equation}

It is reversible Newton equation. The irreversibility in Newtonian
systems can be compatible with reversibility of the Newton
equation in the case if exists the parameter determining
irreversibility, concerning which Newton equation is invariant.
Such parameter is a velocity of the center of mass of subsystem.
The Newton equation for subsystem elements is invariant in respect
to this velocity. But an exchange of energy between subsystems is
not invariant to this velocity. These energy exchanges are
determined by the work of the generalized forces, which are
proportional to the speed of change of the velocity of the center
of mass of subsystem.

Easiest explanation of the fact, why the Newton equation does not
determine the generalized forces, can be made on the example of
system consisting of two balls connected by a spring. Shortly, in
the laboratory system of coordinate the energy of each ball
simultaneously transforms in kinetic energy of other ball and in
the potential spring energy. But the Newton equation describes
only transformation of kinetic energy into the potential energy.
Therefore to solve this task with the help of the Newton equation,
it is necessary to exclude energy of the system motion as the
whole. For two bodies it is possible to do by transition to the
coordinates system of the center of mass. In this coordinate
system the change of kinetic energy of a ball is equally to the
change (but with opposite value) of the potential energy. But for
three and more bodies it is impossible to do. Therefore $N\geq3$
bodies system in general case cannot be reduced to the independent
equations of Newton [4,8-10].

Therefore, for the description of evolution of nonequilibrium
system it is necessary to obtain the equation determining
generalized forces or the speed of change of the velocity of the
center of mass of system, in results of its interaction with other
subsystems. For this purpose we'll do the following. In the some
system of coordinates we shall present kinetic energy of system as
energy of subsystem motion as the whole - $T_n^{tr}$, kinetic
energy of motion of its elements concerning the center of mass
-$\widetilde{T}_N^{ins}$, a potential energy of subsystems
elements- $\widetilde{U}_N^{ins}$ and interaction energy with
other subsystems. The energy
$E_N^{ins}=\widetilde{T}_N^{ins}+\widetilde{U}_N^{ins}$ we'll call
a binding energy. At absence of external forces the energy
$T_N^{tr}$ and $E_N^{ins}$ are the motion integrals. Their sum is
a full energy of the system. We shall assume, that the system is
prepared by nonequilibrium way so, that it can be divided into two
equilibrium subsystems. Then it is possible to generalize results
of their analysis on any nonequilibrium systems.

The following equations for exchange of energy between subsystems
have a place [9]:
\begin{equation}
{LV_L\dot{V}_L+{\sum\limits_{j=i+1}^L}\sum\limits_{i=1}^{L-1}\{v_{ij}
[\frac{\dot{v}_{ij}}{L}+\frac{\partial{U}}{\partial{r_{ij}}}]\}=
-\sum\limits_{{j_K}=1}^K}\sum\limits_{{i_L}=1}^{L}v_{i_L}
\frac{\partial{U}}{\partial{r_{{i}_{L}{j}_{K}}}} \label{eqn8}
\end{equation}

\begin{equation}
{KV_K\dot{V}_K+{\sum\limits_{j=i+1}^K}\sum\limits_{i=L+1}^{K-1}\{v_{ij}
[\frac{\dot{v}_{ij}}{K}+\frac{\partial{U}}{\partial{r_{ij}}}]\}=
-\sum\limits_{{j_K}=1}^K}\sum\limits_{{i_L}=1}^{L}v_{j_K}
\frac{\partial{U}}{\partial{r_{{i}_{L}{j}_{K}}}} \label{eqn9}
\end{equation}
Here we take ${LV_L+KV_K=0}$, ${V_L}$ and ${V_K}$ is a velocities
of the center of mass of $L$ and $K$ subsystems; $L$ and $K$ are
number of elements in subsystems; ${v_{ij}}$ -is a relative
velocities $i$ and $j$ elements; ${L+K=N}$. Masses of elements are
accepted to $1$. Double indexes are accepted for a designation, to
what subsystems these elements belong.

The left side in eq.(8,9) determines exchanges of the energy
between subsystems. The first members set change kinetic energy of
motion of subsystems, ${T^{tr}}$ as the whole. The second members
describe transformation of binding energy of the subsystems,
${E^{ins}}$.

The right side in the equations eq. (8,9) determines potential
energy of interaction of subsystems. They cause transformation of
kinetic energy of the motion of subsystems, ${T^{tr}}$, to their
binding energy.

In case ${L=K}$ the equation of the motion for one subsystem can
be obtained from eq.(8,9) [9]:
\begin{equation}
{\dot{V}_L=- {\frac{1}{{V_L}}L}
{\sum\limits_{j=i+1}^L}\sum\limits_{i=1}^{L-1}\{v_{ij}
[\frac{\dot{v}_{ij}}{L}+\frac{\partial{U}}{\partial{r_{ij}}}]\}
-\sum\limits_{{j_K}=1}^K}\sum\limits_{{i_L}=1}^{L}
\frac{\partial{U}}{\partial{r_{{i}_{L}{j}_{K}}}} \label{eqn10}
\end{equation}

The equation (10) is determining the generalized force. This force
is depending not only from coordinates but from velocities also.
Presence of the dependence from the velocities is a necessary
condition of irreversibility. Therefore we will call this equation
as evolutionary equation. With the help of eq.(2) it is not
difficult to show  that equation (10) is correct for hard-disks
system.

In a limit ${N\gg1}$ for enough fine-grained splitting system, the
equation (10) determines a field of generalized forces between the
subsystems, caused by their relative motion. Irreversibility is
connected with decrease of the generalized forces in a result of
transition of kinetic energy of motion of a subsystem in the
binding energy of subsystems. For any splitting of the equilibrium
system the relative velocities of subsystems and the energy flow
between them are equal to zero [9, 11].

Let us explain, why the generalized forces depend from velocities
of the elements. Inherently, this force determines the change of
velocity of the center of mass. But this velocity is determined by
distribution of elements velocities. Therefore the interaction
force of subsystems should depend not only from the forces of
potential interaction of elements, but also from distribution of
their velocity.

Laws of conservation of energy and momentum do not forbid for
equilibrium system of spontaneous occurrence of an inequality
${\dot{T}^{tr}>0}$ for some of subsystems. This inequality means
infringement of equilibrium [10, 11] and transition from canonical
Hamilton formalism systems description to the description of
system inside the framework of the generalized equations. It is
connected with appearance of the generalized forces between
subsystems. But canonical Liouville equation cannot be transformed
by spontaneous way to the generalized type. It is impossible
because of conservation of the canonical Hamilton's formalism.
From the physical point of view it is impossible because of
stability of an equilibrium state for mixed systems [8]. Such
stability is caused by aspiration to zero of the generalized
forces arising at a deviation of system from equilibrium [8, 11,
and 14]. Hence, the system, having come to equilibrium, never
leaves this condition.

\date{\it{Thermodynamics and classical mechanics.}} \maketitle

Let's consider interrelation of classical mechanics with
thermodynamics. Existence two invariant of the motion-${E^{ins}}$
and ${T^{tr}}$, and also character of their transformation at
interaction of the subsystems, determined by the equation (10),
allow to catch deep analogy between the equations (8-10) and the
basic equation of thermodynamics [14]:
\begin{equation}
{dE=dQ-{PdY}} \label{eqn11}
\end{equation}

Here, according to the common terminology [14], $E$ is internal
energy of a subsystem; $Q$ is a thermal energy; $P$ is a pressure;
$Y$ is a volume.

The change of energy of the selected subsystem is caused by work
of external forces. Therefore to change of full energy of a
subsystem corresponds to $dE$.

The change of kinetic energy of motion of a subsystem as the
whole, $dT^{tr}$, corresponds to the member ${PdY}$. Really,
${dT^{tr}=VdV=V\dot{V}dt=\dot{V}dr=PdY}$

Let's us determine, what the member in the eq.(11) corresponds to
the change of the binding energy of a subsystem. From the virial
theorem [8] it follows that if potential energy is homogeneous
function of the second degree from all radiuses-vectors, then
${\bar{E}^{ins}=2\bar{\tilde{T}}^{ins}=2\bar{\tilde{U}}^{ins}}$.
Feature means averaging on time. Earlier we have obtained, that
the binding energy, ${E^{ins}}$ increases due to the energy,
${T^{tr}}$. But opposite process is impossible. Therefore the
member $Q$ from the eq.(11) is conforming to the energy
${E^{ins}}$.

Let the subsystem will consist of ${N_m}$ elements. Then average
energy of each element, ${\bar{T}_0^{ins}=E^{ins}/N_m}$. Let the
binding energy increases on the value ${dQ}$. In connection with
virial theorem, keeping members of the first order, we shall have:
${dQ\approx\bar{T}_0^{ins}[d\bar{E}^{ins}/\bar{T}_0^{ins}]
=\bar{T}_0^{ins}[{dv}/{v_0}]}$, where ${v_0}$ is average velocity
of an element, and ${dv}$ its change. As the condition the
subsystems equilibrium is accepted, we have
${dv/v_0\sim{{d\Gamma_m}/{\Gamma_m}}}$. Here ${\Gamma_m}$ is a
phase volume of a subsystem, and ${d\Gamma_m}$ is increasing
${\Gamma_m}$ due to increasing of the subsystem energy on the
value, ${dQ}$. By keeping members of the first order we'll have:
${dQ\approx\kappa{\bar{E}^{ins}d\Gamma_m/\Gamma_m=
\kappa{\bar{E}^{ins}}d\ln{\Gamma_m}}}$, where
${\bar{T}_0^{ins}=\kappa{\bar{E}^{ins}}}$. By definition
${d\ln{\Gamma_m}=dS^{ins}}$, where ${S^{ins}}$ is entropy [10,14].
So, we have near equilibrium:
${dQ\approx\bar{T}_0^{ins}dS^{ins}=\kappa\bar{E}^{ins}dS^{ins}}$.

Let's consider the connection of the field of forces of
interaction of subsystems with entropy. As the sum of all forces
of interaction of subsystems at any moment is equal to zero, the
condition, ${\sum\limits_{m=1}^R\dot{T}_m^{tr}=0}$ have a place.
I.e. energy ${T^{tr}}$ is redistributed between subsystems and
goes on increasing of the binding energy. It is equivalent to
entropy increasing. Relative velocities of subsystems go to zero,
and the system equilibrates. Near to equilibrium the right side of
the eq.(5) is equal to zero. So, entropy deviation from
equilibrium state can be determined with the help of following
formula:
\begin{equation}
{{\Delta{S}}={\sum\limits_{l=1}^R{\{{\frac{m_l}{E^{m_l}}}
\sum\limits_{k=1}^{m_l}\int{\sum\limits_s{{F_{ks}}^{m_l}}{dr_k}}\}}}}\label{eqn12}
\end{equation}
Here ${E^{m_l}}$ is a kinetic energy of subsystem; ${m_l}$ is a
number elements in subsystem -${"l"}$; ${R}$- is a number of
subsystems; ${s}$- is external disks which collided with internal
${k}$ disk along trajectory. The integral is determining the work
of the force ${F_{ks}^{m_l}}$ during equilibrating.

The eq.(12) connects the generalized force with entropy. Similar
connection is established by the formula for KS - entropy in which
entropy is expressed through Lyapunov's exponents [3]. Inherently
the eq.(12) corresponds to the formula for entropy:
${S=\sum\limits_{l=1}^R{S_l(E_l^{ins}+T_l^{tr})}}$, (see, [11]).
Really, if ${E_l^{ins}\gg{T_l^{tr}}}$, then
${dS=\sum\limits_{l=1}^R\frac{\partial{S_l}}{\partial{T_l^{tr}}}{dT_l^{tr}}}$
that corresponds to eq.(12).

\date{\it{Summarizing.}} \maketitle

Thus, the dynamics of the open and nonequilibrium systems is
determined by the generalized Liouville equation and the
evolutionary equation of the subsystems motion. The necessary
condition of irreversibility follows from generalized Liouville
equation. This condition is dependence of the generalized forces
from relative velocities of subsystems of any non-equilibrium
system. So, due to the work of the generalized forces, the energy
of relative motion of subsystems is transformed to their binding
energy by irreversibly way and goes on entropy increasing. But the
binding energy cannot be transformed to the energy of relative
motion of subsystems. This is essence of the mechanism of
equilibration [5-9, 11].

Though dynamics of hard disks can be described without use of
potential energy, the essence of the mechanism of equilibration
for them is identical to the mechanism for Newtonian systems.
Difference only in that that in Newtonian systems the energy of
relative motion of subsystems can be transformed to their binding
energy due to the potential interaction of subsystems.

In equilibrium we have ${\dot{T}_L^{tr}}$=${\dot{T}_K^{tr}}$=0 and
${V_L=V_K=0}$, therefore the work of interaction force of each
subsystem, and also energy ${T^{tr}}$ are equal to zero. It is
equivalent to conditions: ${\dot{E}_L^{ins}=\dot{E}_K^{ins}=0}$.
Hence, the closeness of system to equilibrium means the closeness
to zero of the generalized forces between subsystems. The
criterion of equilibrium is applicability of the canonical
Hamilton and Liouville equations for the description of dynamics
of subsystems for any splitting of full system. Only in
equilibrium the dynamics is reversible for both the subsystems and
for the system as a whole. The dynamics is reversible also for the
weak fluctuations because the entropy has a second infinitesimal
order in respect to deviations of a system from equilibrium [9,
11, 13].

Criterion for the sufficiently fine-grained splitting of system is
the closeness of relation ${T^{tr}/{E^{ins}}}$ to zero. It is
obvious that the number of elements in a subsystem should be large
enough. Otherwise fluctuations become commensurable with errors of
a field of generalized forces [11, 13].

Splitting the system onto the equilibrium subsystems is necessary
reception for studying the dynamics laws of the many-body systems.
It allows setting a field of the generalized forces and to
describe it's with the help of the evolutionary equation (7). In
particular, only due to this reception it is possible to find out,
that their work is goes by irreversible way on increase of the
binding energy of a subsystem.

The basic results received here do not contradict results of the
analysis of systems, which was obtained in earlier works by
similar method: splitting into the subsystems and studying its
dynamics near equilibrium. So, in [11] the macroscopical motion
(it is similar to the subsystems motion in our case) was
considered. There by using of the method of multipliers Lagrange
it has been strictly proved that equilibrium is possible only when
${T^{tr}=0}$. Also the validity of the equation for entropy (12)
is follows from it.

Why till now it was impossible to solve a problem of
irreversibility? From our point of view the explanation is
consists in the following. As a rule it was accepted without doubt
that the full description of evolution of nonequilibrium systems
could be realized on the basis of the Newton equation and
canonical Liouville equation. But we find, that the canonical
Liouville and Newton equations are not applicable for describing
work of the generalized forces because in nonequilibrium systems
they are dependent from velocities.

Indeed, let us try to determine the generalized force with the
help of Newton eq.(7) by summation of all forces acting on
elements of a subsystem. Then we obtain the forces independent
from velocities due to invariancy of the Newton equation
concerning Galileo transformation . So, the Newton equation, does
not describe a field of the generalized forces.

Let's to emphasize, that the substantiation of irreversibility
mechanism and connection classical mechanics and thermodynamics
offered here, follows only from laws of conservation of energy and
a momentum.

The evolutionary equation of the motion subsystem along with the
generalized Hamilton and Liouville equations given above, and also
the established relations between classical mechanics and
thermodynamics, can be put into the base of theoretical
description of evolution of the open systems.

\end{document}